\documentclass[12pt,a4paper]{article}

\usepackage[utf8]{inputenc}
\usepackage[T1]{fontenc}
\usepackage[margin=2.5cm]{geometry}
\usepackage[numbered]{bookmark}
\usepackage{mathtools}
\usepackage{amssymb}
\usepackage{graphicx}
\usepackage[font=small,labelfont=bf]{caption}
\usepackage{authblk}
\usepackage{cite}
\usepackage{dsfont}
\usepackage{xcolor}
\usepackage[titles]{tocloft}

\numberwithin{equation}{section}
\newcommand{\eq}[1]{\begin{equation} #1 \end{equation}}

\newcommand{\pa}{\partial}

\newcommand{\cn}{\mathrm{cn}}

\newcommand{\nocontentsline}[3]{}
\newcommand{\tocless}[2]{\bgroup\let\addcontentsline=\nocontentsline#1{#2}\egroup}

\title{Pulsating strings in $Schr_5 \times T^{1,1}$ background}
\author[a,b]{A.~Golubtsova}
\author[a,c]{H.~Dimov}
\author[c]{I.~Iliev}
\author[c]{M.~Radomirov} 
\author[c,d]{R.~C.~Rashkov}
\author[c]{T.~Vetsov \footnote{Emails: \texttt{golubtsova, dimov  @theor.jinr.ru}, \texttt{ivo.n.iliev@abv.bg} \\and \texttt{h\_dimov,radomirov,rash,vetsov@phys.uni-sofia.bg}}}

\affil[a]{\textit{The Bogoliubov Laboratory of Theoretical Physics, JINR,}\authorcr\textit{141980 Dubna, Moscow region, Russia}\vspace{5pt} \vspace{3pt}} 
\affil[b]{\textit{Dubna State University}\authorcr\textit{Universitetskaya str.,141980 Dubna, Moscow region, Russia}\vspace{5pt} \vspace{3pt}} 
\affil[c]{\textit{Department of Physics, Sofia University,}\authorcr\textit{5 J. Bourchier Blvd., 1164 Sofia, Bulgaria}\vspace{5pt} \vspace{3pt}} 
\affil[d]{\textit{Institute for Theoretical Physics, Vienna University of Technology,}\authorcr\textit{Wiedner Hauptstr. 8--10, 1040 Vienna, Austria}}

\date{}
	
\begin{document}
	
\maketitle

\vspace{-20pt}

\begin{abstract}
The quest for extension of holographic correspondence to non-relativistic sectors naturally includes Schr\"odinger backgrounds and their field theory duals. In this paper we study the holography by probing the correspondence with pulsating strings. The case we consider is pulsating strings in five-dimensional Schrödinger space times five-torus $T^{1,1}$, which has as field theory dual a dipole CFT. First we find particular pulsating string solutions and then semi-classically quantize the theory. We obtain the wave function of the problem and thoroughly study the corrections to the energy, which by duality are supposed to give anomalous dimensions of certain operators in the dipole CFT.

\end{abstract}

\tableofcontents
\newpage

\section{Introduction}

In recent years the intensive development of the AdS/CFT correspondence has provided us with powerful tools for studying important aspects of string theory and quantum field theories. This unusual correspondence is a true duality in a sense that the strong coupling quantum regime of one of the theories is equivalent to the weakly coupled semi-classical regime of the other one. The latter enables us to establish a dictionary between objects on both sides of the duality, which opens a window to non-perturbative physics.  
 
Although the correspondence has been successfully applied to the description of highly supersymmetric theories, it has also been extended to less supersymmetric models, which are more interesting for phenomenology. From gravitational point of view, a straight-forward way to reduce supersymmetry in a given model can be achieved by deforming the original spacetime background such as the new background is again an Einstein manifold.

A simple procedure, proposed in  \cite{Lunin:2005jy}, enables us to explicitly generate new supergravity solutions and also gives suggestions about their dual gauge theories. This approach is based on the global symmetries underlying the starting theory. Particularly, having two global $U(1)$ isometries one can interpret them geometrically as a two-torus with action of the associated $SL(2,\mathbb{R})$ symmetry on the torus parameter, which are given by $\tau\to \tilde{\tau}=\tau/(1+\gamma\tau)$. Considering type  IIB backgrounds, one already has $SL(2,\mathbb{R})$ symmetry of the 10-dimensional background. In this case, the existing two $SL(2,\mathbb{R})$ symmetries get combined to form $SL(3,\mathbb{R})$, which can be seen as a compactification of M-theory on $T^3$. However, from holographic point of view, it is important where in the spacetime the two-torus, associated with the global isometries, are located. Indeed, if the torus lives entirely in the asymptotically AdS part of the background geometry the dual gauge theory becomes essentially non-commutative. Here, the $U(1)$ charges are identified as momenta and the product in the field theory is replaced by the Moyal star-product. If however, the torus resides in the complementary part of the given spacetime, the product between the fields is the ordinary product, but the theory exhibits a Leigh-Strassler deformation \cite{Leigh:1995ep}. Since the deformation is taken in a direction along the R-symmetry it is quite clear that the supersymmetry in the new theory will be at least partially broken. This is in agreement with the gauge theory side, as shown in \cite{Leigh:1995ep}.
This solution generating technique can be applied to wide range of cases even with or without the presence of the $SL(2,R)$ symmetry.
A well-known example is the exactly marginal deformations of $\mathcal N = 4$ Super Yang-Mills theory, considered in \cite{Leigh:1995ep}, whose gravity dual (for the so-called $\beta$-deformation) was derived in \cite{Lunin:2005jy}. Other examples also include deformations of Lunin-Maldacena background, which have been thoroughly studied
in the literature, for instance in \cite{Frolov:2005ty,frolovnew, Gursoy:2005cn, Chu:2006ae, Bobev:2006fg, Bobev:2007bm, Bobev:2005cz, Bykov:2008bj, Dimov:2009ut, Frolov:2005dj}. 

Recently, in an attempt to generalize the AdS/CFT correspondence to strongly coupled non-relativistic field theories \cite{Son:2008ye,Balasubramanian:2008dm}, qualitatively different solutions have been obtained. A particularly interesting examples include the non-relativistic Schr\"odinger spacetimes, where the isometry group of the background geometry on string side is the Schr\"odinger group. It consists of time and space translations, space rotations, Galilean boosts and dilatations. It has also been shown that  possible duals to such theories could be realized as the so called dipole gauge theories. These quantum field models are characterized with non-locality, but still living on an ordinary commutative space. Further investigations on dipole theories can be traced for instance in \cite{BG,IKK,oj}. 

One direction, where these solutions have been used, is in the description of ordinary $\mathcal N = 1$ SQCD-like gauge theories, which are considered in the context of D-brane constructions. This type of deformations are typically used  to decouple  spurious effects coming from Kaluza-Klein modes on cycles  wrapped by the D-brane \cite{Gursoy:2005cn, Chu:2006ae, Bobev:2006fg, Bobev:2007bm,Bobev:2005cz}. Furthermore, models with non-relativistic symmetries are related to a number of other physically interesting studies, such as the Sachdev-Ye-Kitaev (SYK) model \cite{Sachdev:1992fk}, Fermi unitary gas \cite{Shin:2008jsa}, and models with trapped supercooled atoms \cite{Son:2008ye,Adams:2008wt}, which in most cases are strongly correlated. 

For this reason, we are motivated to investigate the properties of these holographic models on the string side, where calculations can be performed explicitly. An important step in this line of investigations has been done in \cite{Guica:2017jmq} where strong arguments for integrability and quantitative check of the matching between string and gauge theory predictions have been presented. These studies triggered a new interest to holography in such backgrounds, since they open the option to gain important knowledge about dipole theories for instance. Particular string solutions beyond the supergravity approximation has been found in \cite{Guica:2017jmq,Ahn:2017bio, Roychowdhury:2019olt, Georgiou:2017pvi,Georgiou:2018zkt, Georgiou:2020qnh}. String theory on the pp-wave geometry of the non-supersymmetric Schrödinger background has been studied in \cite{Georgiou:2019lqh}. A semiclassical quantization has been also considered in \cite{Ouyang:2017yko} and very recently finite corrections has been obtained in \cite{Zoakos:2020gyb}.
More aspects of Schr\"odinger holography can be found for instance in \cite{Dimov:2019koi, Golubtsova:2020fpm,Bobev:2009mw, Bobev:2009zf, Blau:2009gd,}. 

%%%
In the current paper we focus on investigating the dynamics of pulsating strings on spacetime with Schr\"odinger symmetry. On the gauge theory side the models are known as non-relativistic duals. Gravity duals of theories with Schr\"odinger symmetry  were obtained in \cite{Balasubramanian:2008dm}. The relevant for our considerations Schr\"odinger part of the background was obtained in \cite{Dimov:2019koi} by TsT deformation along certain directions from $AdS_5\times S^5$. As shown for instance in \cite{Bobev:2009mw}, the generated solutions are twisted and the supersymmetry is completely broken. More detailed  description of this technique can be found for instance in \cite{Lunin:2005jy}. A brief review of pulsating strings in holography has been presented in \cite{Dimov:2019koi}, where the authors also investigated the case of pulsating strings in $Schr_5\times S^5$. Our considerations will be restricted to the bosonic sector of the theory with complementary manifold of type $T^{1,1}$ conifold, which is interesting as being a part of the Calabi-Yau manifolds from the string landscape.  

Our studies begin in Section \ref{sec2}, where we consider the standard Polyakov string action in conformal gauge for pulsating strings in $Schr_5\times T^{1,1}$ spacetime. Here, we impose an appropriate pulsating string anzats and explicitly solve the corresponding classical equations of motion together with the Virasoro constraints in terms of simple trigonometric or Jacobi elliptic functions. Furthermore, in this section we identify the relevant conditions of the parameters for realizing a pulsating string configuration. In Section \ref{sec3} we focus on calculating the energy spectrum, the corresponding first order correction in powers of $\lambda$ for the pulsating string configuration, and the anomalous dimensions of the operators in the dual gauge theory. In this case, the calculations are possible in perturbation theory due to the fact that the squared string Hamiltonian takes the form of a point particle one. Finally, in Section 4 we make a brief conclusion of our results.

%%%%%%%%%%%%%%%%%%%
%\section{Pulsating strings in Schr\"odinger background: solutions}

\section{Pulsating string solutions in $Schr_5 \times T^{1,1}$}\label{sec2}
In this section we will construct the ten-dimensional $Schr_5 \times T^{1,1}$ background by applying a TsT transformation on the original $AdS_5 \times T^{1,1}$, which initially comes without a $B$-field. Consequently, the new background will acquire non-trivial $B$-field, thus essentially changing the character of the string dynamics. However, we show that the classical string equations of motion in $Schr_5 \times T^{1,1}$ admit explicit analytical solutions in therms of simple trigonometric and Jacobi elliptic functions.

\subsection{Generating $Schr_5 \times T^{1,1}$}

It is well know for nearly fifty years that the symmetry of the free Schr\"odinger equation 
\begin{equation}
\frac{\partial^2}{\partial\vec{r}^2}\phi -2im \frac{\partial}{\partial t}\phi =0,
\end{equation}
is the so called Schr\"odinger group. It consists of the following spacetime transformations in $d$ space
dimensions
\begin{equation}
t\:\to\: t'=\frac{at+b}{ct+d}, \qquad \vec{r}\:\to\: \vec{r}\,'=\frac{\Omega\vec{r} + \vec{v}t + \vec{A}}{ct+d}.
\end{equation}
Here $\Omega$ is a rotation matrix and $ad-bc=1$. The scaling properties of spacial directions and the time coordinate are different, $(t,\vec{x})\:\to\:(\lambda^2 t,\lambda \vec{x})$.

From group point of view, the Schr\"odinger group can be thought of as a non-relativistic analogue of the conformal group in $d$ dimensions. In fact, as can be seen from above, the Schrödinger group in $d$ spatial dimensions can be embedded into the relativistic conformal group in $d+1$ dimensions $SO(2,d+2)$, or particular contraction of the conformal group \cite{barut-niederer}. 

For purposes of holographic correspondence it is important to consider spaces with the Schr\"odinger group being the maximal group of isometries.
As it was mentioned in the introduction, there is a simple way to obtain the spacetime geometry possessing this symmetry. Having the comments above, it is a natural guess to assume that it can be done starting from the AdS space associated with the corresponding conformal group. Being particular case of the so called Drinfe'ld-Reshetikhin twist, the TsT procedure has been used for generating many backgrounds keeping integrability (or part of it) intact. Specific point in generating Schr\"odinger backgrounds via TsT transformations is that one of the light-cone variables is involved. This type deformations are known also as null-Melvin twist. 
 To generate null-Melvin twist we implement the following steps:
 
\begin{itemize}
	\item write the theory in light-cone coordinates and identify a Killing direction, say $\psi$,
	\item T-dualize along the Killing direction $\psi$,
	\item boost the geometry in the Killing direction by an amount $\hat{\mu}$, 
	\item T-dualize the geometry back to IIA/IIB along  $\psi$.
\end{itemize}

In order to accomplish the desired result consider $AdS_5\times X^5$, where the metric on $X^5$ is $g_{\alpha\beta}$. Now we perform a null Melvin twist along a Killing vector $\mathcal{K}$ on $X^5$. The result is
\begin{equation}
ds^2_{10}=ds^2_{Schr_5}+ds^2_{X^5},
\end{equation}
where
\begin{equation}
ds^2_{Schr_5}=-\frac{\Omega}{z^4}+\frac{1}{z^2}(-2dvdu+dx_1^2+dx_2^2+dz^2), \quad \Omega= \left\| \mathcal{K}\right\| ^2=g_{\alpha\beta}\mathcal{K}^\alpha \mathcal{K}^\beta.
\end{equation}
It is clear that $\Omega$ is non-negative being a square length of the Killing vector. The generated $B$-field has the form
\begin{equation}
	B_{(2)}=\frac{1}{z^2}\mathcal{K}\wedge du.
\end{equation}
An important remark is that in order to make holographic sense of these solutions one has to impose some conditions. In particular for these to be holographic duals to non-relativistic field theories the light-cone coordinate $v$ should be periodic, $v\sim v +2\pi r_v$ \cite{Son:2008ye,Balasubramanian:2008dm,Adams:2008wt}. The momentum along this compact direction is quantized in units of the inverse radius $r_v^{-1}$ .

Let us translate the above procedure to our case introducing the notations we will use from now on.
We proceed with $AdS_5 \times T^{1,1}$ background, which arises from a stack of $N$ $D$-branes at the tip of the conifold, the conic Calabi-Yau 3-fold whose base space is $T^{1,1}$ \cite{Klebanov:1998hh}. The manifold $T^{1,1}$ is the coset space $(SU(2) \times SU(2))/U(1)$. This supergraviry solution is the dual to the ${\cal{N}} = 1$ supersymmetric Yang-Mills theory.  The metric of $AdS_5 \times T^{1,1}$ is defined as
\begin{equation}\label{t11}
ds^2 = ds_{AdS_5}^2 \,+\, \ell^2 \frac{b}{4} \left[  
\sum_{i=1}^{2} \left( d\theta_i^2 + \sin^2\theta_i\, d\phi_i^2 \right)  +
b \left( d\psi - \sum_{i=1}^{2} \cos\theta_i\, d\phi_i \right)^{\!2}\, \right]\! ,
\end{equation}
where the metric on $AdS_5$ is written in light-cone coordinates by
\begin{equation}
ds^2_{AdS_5}=\ell^2\,\frac{2dx^+dx^-+dx^idx_i +dz^2}{z^2}.
\label{ads-l-c}
\end{equation}
This is clearly an $S^1$ Hopf fibration over $S^2 \times S^2$ with $\psi$ parameterizing the fiber circle which winds over the two base spheres once\footnote{When the Hopf numbers of the $S^1$ bundle over the base 2-spheres are $p$ and $q$ the above generalizes to the manifolds $T^{p,q}$.}. The isometry group of $T^{1,1}$ is
$SU(2) \times SU(2) \times U(1)$ and in particular, it has three commuting Killing vectors: $\partial/\partial \phi_1, \partial/\partial \phi_2, \partial / \partial \psi$.

It is easily seen that the generated metric by doing TsT along $\psi$ and $x^-$ will produce $5d$ Schr\"odinger spacetime times conifold
\begin{equation}\label{t11}
ds^2 = ds_{Schr_5}^2 \,+\, \ell^2 \frac{b}{4} \left[  
\sum_{i=1}^{2} \left( d\theta_i^2 + \sin^2\theta_i\, d\phi_i^2 \right)  +
b \left( d\psi - \sum_{i=1}^{2} \cos\theta_i\, d\phi_i \right)^{\!2}\, \right]\! ,
\end{equation}
where 
\begin{equation}
ds^2_{Schr_5}=\ell^2\left(-\, \frac{\hat{\mu}^2(dx^+)^2}{z^4} + \frac{2dx^+dx^-+dx^idx_i+dz^2}{z^2} \right)\! ,
\label{metric-schro}
\end{equation}
and there is also a generated $B$-field given as in \eqref{B-field}. Now we proceed with choosing an ansatz for our problem.

\subsection{Ansatz and classical equations of motion}

In the previous subsection the metric has not been written in  global coordinates. 
The ten-dimensional line element of the background in global coordinates is given by 
\begin{equation}
ds^2_{Schr_5\times T^{1,1}}=ds^2_{Schr_5}+ds^2_{T^{1,1}},
\end{equation}
where the Schr\"odinger part of the metric in global spacetime coordinates is written by \cite{Dimov:2019koi}
\begin{equation}
\frac{ds^2_{Schr_5}}{\ell^2}=-\left(1+\frac{\hat{\mu}^2}{Z^4}+\frac{\vec{X}^2}{Z^2}\right)dT^2+\frac{2dTdV+d\vec{X}^2+dZ^2}{Z^2},
\end{equation}
and the $T^{1,1}$ part yields
\begin{equation}
\frac{ds^2_{T^{1,1}}}{\ell^2}=\frac{b}{4}\!\left[\sum\limits_{i=1}^2\left(d\theta_i^2+\sin^2\theta_i\,d\phi_i^2\right)
+b\left(d\psi - \sum\limits_{i=1}^2\cos\theta_i\,d\phi_i\right)^{\!\!2}\,\,\right]\!.
\label{metric_T11}
\end{equation}
Here, $0\leq\psi<4\pi,\ 0\leq\theta_i\leq\pi,\ 0\leq\phi_i<2\pi$, and $b=2/3$, although we will keep the notation $b$ \footnote{One notes that for $b=1$ the $S^5$ metric is recovered.}. Also one has $\hat\mu$ is a deformation parameter coming from the TsT transformation.
Furthermore, the application of TST transformations also generates a non-zero  $B$-field, namely  
\begin{equation}\label{B-field}
B_{(2)} =
\ell^2\dfrac{ b\hat{\mu}}{2Z^2} \,dT\wedge\left( d\psi - \sum\limits_{i=1}^2\cos\theta_i\,d\phi_i \right)\!.
\end{equation}	

The Polyakov string action in conformal gauge, with Kalb-Ramond $B$-field is given by
\begin{equation}
S=-\frac{1}{4\pi\alpha'}\int \!d\tau d\sigma \left(\sqrt{-h}\,h^{\alpha\beta} \pa_\alpha X^M \pa_\beta X^N G_{MN} \,-\,\epsilon^{\alpha\beta} \pa_\alpha X^M \pa_\beta X^NB_{MN}\right)\!,
\end{equation}
where $\alpha,\beta=0,1$, $h_{\alpha \beta}={\rm{diag}(-1,1)}$, $M,N=0,\dots, 9$ and $\epsilon^{01}=+1$. Therefore, the Lagrangian in the considered background takes the following explicit form 
\begin{align}
-4\pi\alpha'\mathcal{L}&=G_{TT} (T'^2-\dot{T}^2) +G_{\vec{X}\vec{X}}( \vec{X}'^2-\dot{\vec{X}}^2) +G_{ZZ}(Z'^2 -\dot{Z}^2) \nonumber\\
&+2G_{TV}(T'V'-\dot{T}\dot{V})+G_{\theta_1\theta_1}(\theta_1 '^2-\dot{\theta}_1^2)+G_{\theta_2\theta_2}(\theta_2 '^2-\dot{\theta}_2^2)\nonumber\\
&+G_{\phi_1\phi_1}(\phi_1'^2-\dot{\phi}_1^2)+G_{\phi_2\phi_2}(\phi_2'^2-\dot{\phi}_2^2)+G_{\psi\psi} (\psi'^2-\dot{\psi}^2)\nonumber\\
&+2G_{\phi_1\phi_2}(\phi_1'\phi_2'-\dot{\phi_1}\dot{\phi_2})+2G_{\phi_1\psi}(\phi_1'\psi'-\dot{\phi_1}\dot{\psi})+2G_{\phi_2\psi}(\phi_2'\psi'-\dot{\phi_2}\dot{\psi})\nonumber\\
&+2B_{T\phi_1}(T'\dot{\phi_1}-\dot{T}\phi_1') +2B_{T\phi_2}(T'\dot{\phi_2}-  \dot{T}\phi_2') +2B_{T\psi}(T'\dot{\psi}- \dot{T}\psi'),
\end{align}
where we have used $\dot X=\partial_\tau X$ and $X'=\partial_\sigma X$. 
In order to obtain pulsating string solutions we consider the following  string ansatz $(\kappa>0)$:  
\begin{align}\label{ansatz}
&T =\kappa \tau, \quad V=0, \quad \vec{X}=\vec{0}, \quad Z=const\neq 0, \nonumber \\
&\theta_1=\theta_1(\tau), \quad \theta_2=\theta_2(\tau), \quad \phi_1=m_1\sigma, \quad \phi_2=m_2\sigma, \quad \psi=m_3\sigma.
\end{align}
In this case, the equations of motion for $V,\,\vec{X},\,\phi_{1},\,\phi_{2}$ and $\psi$, are trivially satisfied, while the equations for $T$, $Z$, $\theta_{1}, \,\theta_{2}$ stay relevant. Let us begin with the equation along $T$, 
\begin{equation}\label{EOMT}
m_1\sin\theta_1(\tau)\,\dot{\theta_1}(\tau)+m_2\sin\theta_2(\tau)\,\dot{\theta_2}(\tau)\,=0.
\end{equation}
It can be written in the following useful form
\begin{equation}\label{eqT1}
m_1\cos\theta_1(\tau)+m_2\cos\theta_2(\tau)\,=A=const,\qquad m_{1,2}\neq 0.
\end{equation}
The equation along $Z$ is given by
\begin{equation}\label{EOMZ}
Z^2=\frac{2\hat\mu \kappa} {b(m_3-A)}=const,
\end{equation}
while the equations along $\theta_i$, $i=1,2$, yield
\begin{equation}\label{EOMTheta12}
\ddot{\theta_i}(\tau)+m_i \sin\theta_i(\tau)\left(m_i \cos\theta_i(\tau)+b(m_3-A) -\frac{2\hat\mu \kappa}{Z^2}\right)=0.
\end{equation}
Using Eq. \eqref{EOMZ} one can write the previous expression in the form
\begin{equation}\label{EOMTHETA}
\ddot{\theta_i}(\tau)+m_i^2 \sin\theta_i(\tau) \cos\theta_i(\tau)=0.
\end{equation}

We should also supplement the equations of motion with the Virasoro constraints:
\begin{align}
&\text{Vir}_1\!:\quad G_{MN}\left(\dot{X}^M\dot{X}^N+X'^MX'^N\right)=0, \label{vir1}\\
&\text{Vir}_2\!:\quad G_{MN}\,\dot{X}^MX'^N=0,\label{vir2}
\end{align}
where the first equation \eqref{vir1} explicitly reads
\begin{align}\label{Vir}
\dot{\theta_1}^2(\tau) + \dot{\theta_2}^2(\tau) + m_1^2\sin^2\theta_1(\tau) + m_2^2\sin^2\theta_2(\tau) - \frac{4\kappa^2}{b}=0,
\end{align}
where we have identified
\begin{equation}\label{key}
\frac{4|G_{TT}| \kappa^2 - b^2(m_3-A)^2}{b}=\frac{4\kappa^2}{b},
\end{equation}
and the second Virasoro constraint \eqref{vir2} is trivially satisfied.
Multiplying \eqref{EOMTHETA} by $\dot{\theta_i}$ we obtain
\begin{equation}
\dot{\theta_i}(\tau)\ddot{\theta_i}(\tau)+ m_i^2\sin\theta_i(\tau) \cos\theta_i(\tau) \,\dot{\theta_i}(\tau)=0,
\end{equation}
thus one finds
\begin{equation}
\frac{d}{d\tau}\left\{\dot{\theta_i}^2(\tau) + m_i^2 \sin^2\theta_i(\tau)\right\}=0,
\end{equation}
or 
\begin{equation}\label{eq_theta}
\dot{\theta_i}^2(\tau) + m_i^2 \sin^2\theta_i(\tau)=K_i>0\,.
\end{equation}
Combining Eqs. \eqref{Vir} and \eqref{eq_theta} we get a relation between the constants $K_1$ and $K_2$, namely
\begin{equation}\label{periods}
K_1+K_2=\frac{4\kappa^2}{b}.
\end{equation}
Now we are in a position to integrate Eq. \eqref{eq_theta}, namely
\begin{equation}
\int\limits_{0}^{\theta_i(\tau)}\frac{d\theta_i}{\sqrt{1-\dfrac{m_i^2}{K_i}\sin^2\theta_i}}=F\left(\theta_i,\sqrt{\frac{m_i^2}{K_i}} \,\right)=\sqrt{K_i}\int\limits_{0}^{\tau} \!d\tau.
\end{equation}
Finally, the solution is given in terms of the Jacobi elliptic cosine
\begin{align}
\label{class_sol1}
\cos\theta_1(\tau)= \cn\left(\!\sqrt{K_1}\,\tau,\sqrt{\frac{m_1^2}{K_1}} \,\right)\!, \qquad
\cos\theta_2(\tau)= \cn\left(\!\sqrt{K_2}\,\tau,\sqrt{\frac{m_2^2}{K_2}} \,\right)\!.
\end{align}
Now the pulsating string condition \eqref{eqT1} acquires the form
\begin{equation}\label{new_puls}
m_1\, \cn\left(\!\sqrt{K_1}\,\tau,\sqrt{\frac{m_1^2}{K_1}} \,\right) + m_2\, \cn\left(\!\sqrt{K_2}\,\tau,\sqrt{\frac{m_2^2}{K_2}} \,\right)=A.
\end{equation}
It must be satisfied for any value of $\tau$. Setting $\tau=0$ we get
\begin{equation}\label{A}
m_1+m_2=A.
\end{equation}

Finally we can find an explicit expression for the classical energy of the considered sting configuration, namely
\begin{equation}
E=-\int\limits_{l/2}^{l/2}\! d\sigma\, \frac{\pa \mathcal{L}}{\pa (\pa_{\tau} T)} = \frac{\ell}{\alpha'} \left[ \left( 1+\frac{\hat{\mu}^2}{Z^4} \right)\kappa + \frac{b\hat{\mu}}{2Z^2}\left( m_1+m_2-m_3 \right) \right], 
\end{equation}
where $l=2\pi\ell$ is the string length and we have also used \eqref{new_puls} and \eqref{A}.

%%%%%%%%%%%%%%%%%%%%%%%%%%%%%%%%%%%%%%%%
\section{Energy corrections and anomalous dimensions}\label{sec3}
In this section we will quantize the pulsating string semi-classically. We will calculate the corresponding energy spectra and their perturbative quantum corrections up to first order in the small parameter $\lambda$. According to the AdS/CFT dictionary, the anomalous dimensions of the corresponding SYM operators are directly related to the corrections of the string energy.

\subsection{Derivation of the Schr\"odinger equation}

In order to semi-classically quantize the bosonic sector of the pulsating string configuration one invokes the Nambu-Goto string action, namely  
\begin{equation} \label{NG}
S_{NG}= -\,\dfrac{1}{2\pi\alpha'} \int \!d\tau d\sigma \,\sqrt{- \det \left(G_{MN}\,\partial_\alpha X^{M}\,\partial_\beta X^{N} - B_{MN}\,\partial_\alpha X^{M}\,\partial_\beta X^{N}   \right) }\,.
\end{equation}

The first ingredient towards finding the spectrum is to make a pullback of the line element of the metric of  $Schr_5 \times T^{1,1}$ to the subspace, where string dynamics takes place. The result for the metric is
\begin{equation}\label{pulsMetric}
ds^2 =\ell^2 \left(-\left|G_{00} \right| dT^2+ \sum\limits_{i,j=1}^2 G_{ij} \,d\theta^i d\theta^j + \sum\limits_{k,h=1}^3 \hat{G}_{kh}(\theta_1,\theta_2) \,d\phi^k d\phi^h \right)\!, 
\end{equation}
where the following quantities have been defined\footnote{From now on we will use the notation $|G_{00}|=|G_{TT}|$.} 
\eq{ 
	\left|G_{00} \right|= 1+ \dfrac{\hat{\mu}^2}{Z^4},\qquad
	\left( G_{ij} \right) =
	\frac{b}{4}\left(\begin{matrix}
		1 & 0 \\
		0 & 1
	\end{matrix}\right)\!,
}
and 
\eq{
	\left(\hat{G}_{kh}\right)=\dfrac{b}{4}\left(\begin{matrix}
		b\cos^2\theta_1+\sin^2\theta_1 & b\cos\theta_1\cos\theta_2 & -b\cos\theta_1\\
		b\cos\theta_1 \cos\theta_2 & b\cos^2\theta_2+\sin^2\theta_2 & -b\cos\theta_2\\
		-b\cos\theta_1 & -b\cos\theta_2 & b
	\end{matrix}\right)\!.
}
The $B$-field also changes to
\begin{equation}\label{B-field}
B_{(2)} =\ell^2 \dfrac{b\hat{\mu}}{2 Z^2} dT\wedge\left( d\psi -\cos\theta_1 d\phi_1 -\cos\theta_2 d\phi_2 \right)= \ell^2\sum\limits_{k=1} ^3 b_{0k} (\theta_1, \theta_2)\,dT\wedge d\phi^k,
\end{equation}
where 
\begin{equation}
b_{01}=-\,\dfrac{b\hat{\mu}}{2Z^2}\cos\theta_1\,,  \qquad   b_{02}=-\,\dfrac{b\hat{\mu}}{2Z^2}\cos\theta_2\, , \qquad b_{03}=\dfrac{b\hat{\mu}}{2Z^2}.
\label{notation-hamiltonian}
\end{equation}
Based on \cite{Arnaudov:2010dk}, in this section we can consider a more extended ansatz than \eqref{ansatz}, such as 
\begin{align}\label{new ansatz}
&T =\kappa \tau, \quad V=0, \quad \vec{X}=\vec{0}, \quad Z=const\neq 0, \nonumber \\
&\theta_1=\theta_1(\tau)\,, \quad \theta_2=\theta_2(\tau)\,, \\
&\phi_1=m_1\sigma+h_1(\tau)\,, \quad \phi_2=m_2\sigma+h_2(\tau)\,, \quad \psi=\phi_3=m_3\sigma+h_3(\tau).
\end{align} 
Taking into account the above ansatz one can also calculate the components of the induced metric and the  $B$-field on the worldsheet: 
\begin{align}
&\frac{ds^2 _{ws}}{\ell^2}=  \left(-|G_{00}|\kappa^2 + G_{ij}\,\dot{\theta}^i\dot{\theta}^j + \hat{G}_{pq}\,\dot{h}^p\dot{h}^q \right) d\tau^2 + \hat{G}_{pq}\,m_p m_q\, d\sigma^2 +
2\hat{G}_{pq}\, m_p\dot{h}^q\, d\tau d\sigma  , \label{ws-m}\\
&B_{(2)}^{ws}=\ell^2 B_{\tau\sigma}\,d\tau\wedge d\sigma ,\qquad \quad B_{\tau\sigma}=-B_{\sigma\tau}=\sum\limits_{i=1} ^3 b_{0i} (\theta_1, \theta_2)\,\kappa\,m_i  \label{ws-b}.
\end{align}
By taking into account \eqref{notation-hamiltonian} one finds that $B^2_{\tau\sigma}\,\equiv\,B^2 (\theta_1,\,\theta_2)$, i.e.
\begin{equation}
B_{\tau\sigma}^2 \,\equiv\,B^2 (\theta_1,\theta_2)\,=\, \frac{b^2{\hat{\mu}}^2\kappa^2}{4\,Z^4}\,\left( m_3 -m_1\,\cos\theta_1 -m_2\,\cos\theta_2 \right) ^2\,\geq\,0\,.
\end{equation}
Now, we can write the Nambu-Goto action \eqref{NG} in terms of the notations we have introduced above,
\begin{equation} \label{NG-ws}
S_{NG}= -\,\dfrac{\ell^2}{\alpha^\prime }\int\! d\tau\, \sqrt{ \left\|\vec{m}\right\|^2 \left( |G_{00}| \kappa^2 - G_{ij}\,\dot{\theta}^i \dot{\theta}^j - \hat{G}_{pq}\,\dot{h}^p\dot{h}^q  \right) + \left( \hat{G}_{pq}\, m_p\dot{h}^q \right)^2  - B^2 }\,,
\end{equation}
where $\ell^2/\alpha ^ \prime =\sqrt{\lambda}$ is the 't Hooft coupling constant and 
\begin{equation}\label{A2}
\left\|\vec{m}\right\|^2 =
\sum\limits_{k,h=1}^3 \hat{G}_{kh}(\theta_1,\theta_2) m_k m_h =\dfrac{b}{4}\left[ \sum\limits_{i=1}^2 m_i^2 \sin ^2\theta_i + b\left( m_3 - \sum_{i=1}^{2} m_i \cos\theta_i \right)^{\!\!2}\, \right] >0,
\end{equation}
or equivalently
\begin{align}
\left\|\vec{m}\right\|^2 &= \frac{b}{4} \left\lbrace \left( b\cos^2\theta_1 + \sin^2\theta_1 \right)\! m_1^2  + \left( b\cos^2\theta_2 + \sin^2\theta_2 \right)\! m_2^2 + b m_3^2 \right. \nonumber\\
&\qquad\,\,\, \left. +2b\cos\theta_1 \cos\theta_2\, m_1 m_2 -2b\cos\theta_1\, m_1m_3 - 2b\cos\theta_2\, m_2m_3 \right\rbrace .
\end{align}

It is useful to consider the  Hamiltonian formulation of the problem. In our case, the canonical momenta are given by
\begin{align}\label{momenta}
\Pi_i&=\frac{\pa L}{\pa \dot{\theta}^i} =\sqrt{\lambda} \,\dfrac{ \left\|\vec{m}\right\|^2 G_{ij}\,\dot{\theta}^j } {\sqrt{ \left\|\vec{m}\right\|^2 \left( |G_{00}| \kappa^2 - G_{kj}\,\dot{\theta}^k \dot{\theta}^j - \hat{G}_{pq}\,\dot{h}^p\dot{h}^q  \right) + \left( \hat{G}_{pq}\, m_p\dot{h}^q \right)^2 - B^2 }} \,, \\
\hat{\Pi}_p&=\frac{\pa L}{\pa \dot{h}^p} =\sqrt{\lambda}\, \dfrac{ \left\|\vec{m}\right\|^2 \hat{G}_{pq}\, \dot{h}^q - (\hat{G}_{kq}\, m_k\dot{h}^q)\, \hat{G}_{pq}\, m_q }
{\sqrt{ \left\|\vec{m}\right\|^2 \left( |G_{00}| \kappa^2 - G_{ij}\,\dot{\theta}^i \dot{\theta}^j - \hat{G}_{kq}\,\dot{h}^k\dot{h}^q  \right) + \left( \hat{G}_{kq}\, m_k\dot{h}^q \right)^2 - B^2 }}\,,
\end{align}
which also implies the constraint 
\begin{equation}
\sum_{p=1}^{3} m_p\, \hat{\Pi}_p=0.
\end{equation}
Using Legendre transformation, $L=\Pi_k\,\dot{\theta}^k-H$, we find the (square of) the pulsating string Hamiltonian
\begin{align}\label{Hamiltonian1}
H^2 = \frac{\left\| \vec{m}\right\|^2 \left|{G_{00}}\right|\kappa ^2 - B^2}{ \left\|\vec{m}\right\|^2} \left(\sum\limits_{i,j=1}^2 G^{ij}\, \Pi_i\Pi_j + \sum\limits_{p,q=1}^3 \hat{G}^{pq}\, \hat\Pi _p \hat\Pi_q \right) + \lambda \left( \left\|\vec{m}\right\|^2 |G_{00}| \kappa^2 - B^2 \right).
\end{align}
We observe that $H^2$ looks like a point-particle Hamiltonian, which seems to be characteristic feature of pulsating strings in holography. The last term, 
\begin{align}\label{potential}
U(\theta_1,\theta_2)=&\, \left\| \vec{m}\right\|^2 \left|G_{00} \right| \kappa^2 - B^2 \nonumber\\
=&\,\frac{b\kappa^2}{4}\left\lbrace b\left( m_1^2+m_2^2+m_3^2 \right) + \left( \left|G_{00}\right| -b\right)\! \left[ m_1^2\sin^2\!\theta_1 + m_2^2\sin^2\!\theta_2 \right]  \right. \nonumber\\
&\left. \qquad\quad +\, 2b\, m_1m_2 \cos\theta_1 \cos\theta_2  - 2bm_3 \left( m_1\cos\theta_1 + m_2\cos\theta_2 \right)  \right\rbrace ,
\end{align}
serves as an effective potential, which encodes the relevant dynamics of the strings.
The $H^2$ is an effective point-particle  Hamiltonian, which can write in the following form
\begin{equation}\label{HamiltonianPoint}
H^2 = \frac{ \left\|\vec{m}\right\|^2 |G_{00}| \kappa^2 - B^2}{ \left\|\vec{m}\right\|^2 } \,\vec{P}^2 + \lambda \,U(\theta_1,\theta_2)\,.
\end{equation}
It is clear that
\begin{equation}
\frac{ \left\|\vec{m}\right\|^2 |G_{00}| \kappa^2 - B^2 }{ \left\|\vec{m}\right\|^2 }\,>\, 0\,.
\end{equation}
The kinetic term of the Hamiltonian \eqref{Hamiltonian1} is considered as a five dimensional Laplace-Beltrami operator of $\,T^{1,1}\,$
\begin{equation}
\vec{P}^2\,=\left(\sum\limits_{i,j=1}^2 G^{ij}\, \Pi_i\Pi_j + \sum\limits_{p,q=1}^3 \hat{G}^{pq}\, \hat\Pi _p \hat\Pi_q \right)  \,\,\longrightarrow \,\, \Delta_{T^{1,1}}\,,
\end{equation}
which defines the eigen-functions of the Hamiltonian, satisfying the following Schr\"odinger equation
\begin{equation}\label{shcrEq}
\frac{ \left\|\vec{m}\right\|^2 |G_{00}| \kappa^2 - B^2 }{ \left\|\vec{m}\right\|^2 } \,\Delta_{T^{1,1}} \,\Psi\,= \,-\,E^2 \,\Psi ,
\end{equation}
or
\begin{equation}\label{Schro}
\left[ 1-\frac{ B^2 (\theta_1,\theta_2) }{ |G_{00}| \kappa^2\, \left\|\vec{m}\right\|^2\! (\theta_1,\theta_2) } \right]  \Delta_{T^{1,1}} \,\Psi\,= \,-\,\frac{E^2}{|G_{00}| \kappa^2}\,\Psi.
\end{equation}
Here we have
\begin{equation}\label{coefficient}
0\,\leq\, \frac{ B^2 (\theta_1, \theta_2) }{ |G_{00}| \kappa^2\, \left\|\vec{m}\right\|^2\! (\theta_1,\theta_2) } \,=\,\frac{ b\,\hat{\mu}^2 }{ Z^4 + \hat{\mu}^2 }\,\frac{ \left( m_3 -m_1\,\cos\theta_1 - m_2\,\cos\theta_2 \right)^2  }{\left[ \sum\limits_{i=1}^2 m_i^2 \sin ^2\theta_i + b\left( m_3 - \sum\limits_{i=1}^{2} m_i \cos\theta_i \right)^{\!2}\, \right] } <1.
\end{equation}
Moreover, at high energies, the inequality is satisfied
\begin{equation}\label{coefficient1}
0\,\leq\, \frac{ B^2 (\theta_1, \theta_2) }{ |G_{00}| \kappa^2\, \left\|\vec{m}\right\|^2 (\theta_1,\theta_2) } \,\ll\,1\,.
\end{equation}

At quantum level the wave function accounts for all quantum fluctuations. As far as we are interested in the quasi-classical quantum behavior it is convenient to split the fluctuations into fast and slow variables. At the relevant accuracy level the wave function ``feels''  only the slow fluctuations while the fast ones enter the wave equation with their average over the fluctuation period. In our case fast variables are the angles $\theta_1$ and $\theta_2$. Indeed, it is easy to see from the explicit solutions \eqref{class_sol1} and \eqref{periods} that the periods $m_i/\sqrt{K_i}$ of $\theta_i(\sqrt{K_i}\tau)$ are very small provided  $\kappa^2$ is very large.
In this sense, the wave function will ``feel'' the average values of
 $\cos\theta_i$ and $\cos^2\theta_i$. Moreover, their oscillation
 amplitudes are in term with a weight much smaller than one \eqref{coefficient1}.
Therefore, we can approximate the small term \eqref{coefficient1} as follows
%%%%%%%%%%%%%%%%%%%%%%%%%%%%%%%%%%%%%
%
\begin{equation}\label{approx coefficient}
0\,\leq\,\frac{ B^2(\theta_1,\theta_2) }{ |G_{00}| \kappa^2\, \left\|\vec{m}\right\|^2 (\theta_1,\theta_2) } \,\approx\, \frac{ b\,\hat{\mu}^2 }{ (Z^4 + \hat{\mu}^2) }\,\frac{ \left( m_3^2 + \frac{1}{2}m_1^2 + \frac{1}{2}m_2^2 \right) }{ \left( b\, m_3^2 + \frac{1+b}{2}\,m_1^2 + \frac{1+b}{2} \,m_2^2 \right)}\,\ll\,1\,.
\end{equation}
Then, we have to solve the following equations
\begin{equation}\label{approx Schrodinger}
\Delta_{T^{1,1}} \,\Psi\,= \,-\,\frac{E^2}{\kappa^2}\,\frac{Z^4\,\left[ 2b\,m_3^2 + (1+b)(m_1^2 + m_2^2) \right] }{ Z^4 \left[2b\,m_3^2 + (1+b)(m_1^2 + m_2^2) \right] + \hat{\mu}^2 \left(m_1^2 +m_2^2 \right)  } \,\,\Psi\,.
\end{equation}
%
%
%
%%%%%%%%%%%%%%%%%%%%%%%%%%%%%%%%%%%%%%%%%%%%%%%%%%%%%%%%%%%%%%%%%%%%%%%%%%%%%%%%%%%%%%%%%%%%%%%%%%%%%%%%%%%%%%%%%%%%%%%%%%%%%%%%%%%%%%%%%%%%%%%%%%%%%%%%%%%%%%%%%%%%%%%%%%%%%%%
\subsection{Laplace-Beltrami operator and wave function}
%%%%%%%%%%%%%%%%%%%%%%%%%%%%%%%%%%%%%%%%%%%%%%%%%%%%%%%%%%%%%%%%%%%%
\paragraph{Laplace-Beltrami operator on $T^{1,1}.$}

Using the line element of $T^{1,1}$ given by \eqref{metric_T11} and the standard definition of the Laplace-Beltrami operator we find (see section 3 of \cite{Gubser:1998vd} for the general case of $T^{p,q}$)
\begin{align}
&\Delta_{T^{1,1}}=\dfrac{4}{b^2}\!\left[\dfrac{b}{\sin\theta_1}\dfrac{\pa}{\pa\theta_1}\!\left(\!\sin\theta_1\dfrac{\pa}{\pa\theta_1}\!\right)\!+
\dfrac{b}{\sin^2\theta_1}\dfrac{\pa^2}{\pa\phi_1^2}+\dfrac{2b\cos\theta_1}{\sin^2\theta_1}\dfrac{\pa^2}{\pa\phi_1\pa\psi}\right.\\
&\left.+\,\dfrac{b}{\sin\theta_2}\dfrac{\pa}{\pa\theta_2}\!\left(\!\sin\theta_2\dfrac{\pa}{\pa\theta_2}\!\right)\!+
\dfrac{b}{\sin^2\theta_2}\dfrac{\pa^2}{\pa\phi_2^2}+\dfrac{2b\cos\theta_2}{\sin^2\theta_2}\dfrac{\pa^2}{\pa\phi_2\pa\psi}+
\!\left(\!1+\dfrac{b\cos^2\theta_1}{\sin^2\theta_1}+\dfrac{b\cos^2\theta_2}{\sin^2\theta_2}\!\right)\!\dfrac{\pa^2}{\pa\psi^2}\right]\!.
\nonumber
\end{align}
We can rewrite this expression in a form, which is more useful for our analysis
\begin{align}\label{laplas}
\Delta_{T^{1,1}}&=\dfrac{4}{b^2}\!\left\lbrace
b\!\left[\dfrac{1}{\sin\theta_1}\dfrac{\pa}{\pa\theta_1}\!\left(\sin\theta_1\dfrac{\pa}{\pa\theta_1}\right)\!+
\dfrac{1}{\sin^2\theta_1}\!\left(\dfrac{\pa}{\pa\phi_1}+\cos\theta_1\dfrac{\pa}{\pa\psi}\right)^{\!\!2}\,\right]\right. \nonumber\\	&\left.+\,b\!\left[\dfrac{1}{\sin\theta_2}\dfrac{\pa}{\pa\theta_2}\!\left(\sin\theta_2\dfrac{\pa}{\pa\theta_2}\right)\!+
\dfrac{1}{\sin^2\theta_2}\!\left(\dfrac{\pa}{\pa\phi_2}+\cos\theta_2\dfrac{\pa}{\pa\psi}\right)^{\!\!2}\,\right]\!+
\dfrac{\pa^2}{\pa\psi^2}\right\rbrace.
\end{align}
The full measure on $T^{1,1}$ is
\begin{equation}
d\Omega=\sqrt{{\rm det}(G_{\mu\nu})}\, d\theta_1 d\theta_2\, d\phi_1 d\phi_2\, d\psi =\frac{b^3}{32}\, \sin\theta_1 \sin\theta_2\, d\theta_1 d\theta_2\, d\phi_1 d\phi_2\, d\psi\,.
\label{measure1}
\end{equation}

\paragraph{Wave function.}

The Schr\"{o}dinger equation for the wave function is
\begin{equation}
\Delta_{T^{1,1}}\,\Psi(\theta_1,\theta_2,\phi_1,\phi_2,\psi)\,=\,-\,M^2\,\frac{E^2}{\kappa^2}\,\Psi(\theta_1,\theta_2,\phi_1,\phi_2,\psi)\,,
\label{schro}
\end{equation}
where 
\begin{equation}
M^2\,=\,\frac{Z^4\,\left[ 2b\,m_3^2 + (1+b)(m_1^2 + m_2^2) \right] }{ Z^4 \left[2b\,m_3^2 + (1+b)(m_1^2 + m_2^2) \right] + \hat{\mu}^2 \left(m_1^2 +m_2^2 \right) } \,.
\end{equation}
One notes that the order of magnitude for $M^2 \sim 1$, when we are situated far from the boundary, $Z\gg 1$.  
To separate the variables, we define $\Psi$ as
\begin{equation}\label{wave}
\Psi(\theta_1,\theta_2,\phi_1,\phi_2,\psi) = e^{il_1\phi_1}\, e^{il_2\phi_2}\, e^{il_3\psi}\, f_1(\theta_1)\, f_2(\theta_2), \qquad
l_1,\,l_2,\,l_3\, \in\mathbb{Z}\,.
\end{equation}
With this choice we can solve for the eigenfunctions, replacing the derivatives along Killing directions $(\partial_{\phi_1},\partial_{\phi_2},\partial_\psi)$ by $(il_1,il_2,il_3)$ correspondingly. 
Substituting \eqref{wave} in \eqref{schro}, together with \eqref{laplas}, we arrive at
\begin{equation}
E^2=\dfrac{4}{b^2}(bE_1^2+bE_2^2+l_3^2)\,,
\label{energy^2}
\end{equation}
where $E_1$ and $E_2$ are determined by the ordinary differential equations
\begin{equation}
\left[\dfrac{1}{\sin\theta_i}\dfrac{d}{d\theta_i}\!\left(\sin\theta_i\dfrac{d}{d\theta_i}\right)\!-
\dfrac{1}{\sin^2\theta_i}\left(l_i+\cos\theta_i l_3\right)^2\right]\!f_i(\theta_i)=\,-\,M^2\,\frac{E_i^2}{\kappa^2}\,\,f_i(\theta_i)\,,\quad i=1,2\,.
\end{equation}
It is convenient to define new variables $z_i=\cos\theta_i$. Then the equations can be written as
\begin{equation}
\left((1-z_i^2)\,\dfrac{d^2}{dz_i^2}-2z_i\,\dfrac{d}{dz_i}-\dfrac{1}{1-z_i^2}\,(l_i+z_i\,l_3)^2 \,+\, M^2\, \frac{E_i^2}{\kappa^2}    \right)f_i(z_i)=0\,.
\end{equation}
The solutions to these equations are given by the standard hypergeometric functions. However, we choose the regular solution
\begin{multline}
f_i(z_i) = (1-z_i)^{\frac{\beta_i}{2}}\,(1+z_i)^{\frac{\alpha_i}{2}}\, {_2F_1}\!\left[\dfrac{1}{2}\!\left( \alpha_i + \beta_i +\,1- \sqrt{1+4\left( l_3^2 + M^2\frac{E_i^2}{\kappa^2}\right) }\, \right)\!,\right.\\
\left.\dfrac{1}{2}\!\left( \alpha_i + \beta_i + 1 + \sqrt{1+4\left( l_3^2 + M^2\frac{E_i^2}{\kappa^2}\right) } \right)\!, \,1\,+ \alpha_i \, ,\,\dfrac{1+z_i}{2}\right]\!,
\end{multline}
where $\alpha_i\equiv\ \mid\!l_i-l_3\!\mid$ and $\beta_i\equiv\ \mid\!l_i+l_3\!\mid$.
In addition, we have to ensure that the solutions $f_i(\theta_i)$ are square integrable with respect to the measure for $\theta_i$, which leads to the following restrictions on the parameters
\begin{equation}
\sqrt{1+4\left( l_3^2 + M^2\frac{E_i^2}{\kappa^2}\right) } - \alpha_i - \beta_i - 1=\,2n_i\,,\qquad n_i\in\mathbb{N}\,.
\label{enpar}
\end{equation}
From \eqref{energy^2} and \eqref{enpar} one finds the squared semi-classically quantized energy of the pulsating string configuration
\begin{equation}\label{Energy}
E^2=\frac{4\kappa^2}{b^2M^2}\!\left(\frac{b}{4}\sum_{i=1}^2(2n_i + \alpha_i + \beta_i +1)^2 -\left( 2b - \frac{M^2}{\kappa^2} \right)\! l_3^2-\frac{b}{2}\right)\!.
\end{equation}
Now, the solution can be written in terms of Jacobi polynomials \cite{Braaksma_1968}
\begin{equation}\label{solution}
f_i(z_i)=(1-z_i)^{\alpha_i/2}\,(1+z_i)^{\beta_i/2}\,\dfrac{n_i!\,\Gamma(\alpha_i+1)}{\Gamma(\alpha_i+1+n_i)}P^{(\alpha_i,\beta_i)}_{n_i}(z_i)\,,
\end{equation}
thus one can write the normalized wave functions such as
\begin{align}
\Psi_{n_i}^{\alpha_i,\beta_i}(z_i)=&\,\sqrt{\frac{(\alpha_i+\beta_i+1+2n_i)\,n_i!\,\Gamma(\alpha_i+\beta_i+1+n_i)}{2^{\alpha_i+\beta_i+1}\,
\Gamma(\alpha_i+1+n_i)\,\Gamma(\beta_i+1+n_i)}} \nonumber\\
&\times(1-z_i)^{\alpha_i/2}\,(1+z_i)^{\beta_i/2}P^{(\alpha_i,\beta_i)}_{n_i}(z_i)\,,\qquad i=1,2\,.
\label{wfun}
\end{align}
Therefore, the full wave function in $T^{1,1}$ \eqref{wave} looks like
\begin{equation}
\Psi=\frac{1}{\sqrt{16\pi^3}}\, e^{il_1\phi_1}\, e^{il_2\phi_2}\, e^{il_3\psi}\,\,\Psi_{n_1}^ {\alpha_1,\beta_1}(z_1)\, \Psi_{n_2}^{\alpha_2,\beta_2}(z_2).
\end{equation}
Taking the limit $\hat{\mu}\,\rightarrow\,0$  ($ Schr_5 \times T^{1,1}\, \rightarrow\, AdS_5 \times T^{1,1}\,$)  we find perfect agreement with the results for the energy and the wave function given in \cite{Arnaudov:2010dk}.

\subsection{Leading correction to the energy}
%%%%%%%%%%%%%%%%%%%%%%%%%%%%%%%%%%%%%%%%%%%%%%%%%%%%%%%%%%%%%

In terms of the new variables $z_i$ the potential \eqref{potential} becomes
\begin{align} \label{potential}
U&(z_1,z_2)= \left\| \vec{m}\right\|^2 |G_{00}| \kappa^2 - B^2 \nonumber\\
&=\frac{b\kappa^2}{4}\left\lbrace b\left( m_1^2+m_2^2+m_3^2 \right) + \left( |G_{00}| -b \right)\! \left[ m_1^2 \left( 1-z_1^2 \right)  + m_2^2 \left( 1-z_2^2 \right) \right]  \right. \nonumber\\
&\left. \qquad\qquad +\, 2b\, m_1m_2\, z_1z_2 - 2b m_3 \left(m_1z_1+m_2z_2\right)  \right\rbrace .
\end{align}
Consequently the measure \eqref{measure1} also changes to
\begin{equation}
d\Omega= \dfrac{b^3}{32}\, dz_1\, dz_2\, d\phi_1\, d\phi_2\, d\psi \,,  \quad  -1\leq z_1,z_2\leq1\,.
\end{equation}
Hence, the first correction to the energy is given by the expression
\begin{align}\label{delta-E}
\delta E^2
&=\lambda \int\limits_{\!\!\!\!\!-1}^1\! \int\limits_{\!\!\!\!-1}^1\! \int\limits_0^{2\pi}\! \int\limits_0^{2\pi}\! \int\limits_0^{4\pi}\, \left| \Psi(z_1, z_2, \phi_1,\phi_2,\psi) \right|^2\,\, U(z_1,z_2)\,\, d\Omega(z_1, z_2, \phi_1,\phi_2,\psi) \nonumber\\
&=\lambda\, \frac{b^3}{32} \int\limits_{\!\!\!\!\!-1}^1\! \int\limits_{\!\!\!\!-1}^1  \left|\Psi_{n_1}^{\alpha_1,\beta_1}(z_1)\right|^2 \left| \Psi_{n_2}^{\alpha_2,\beta_2}(z_2)\right|^2\, U(z_1,z_2)\, dz_1\, dz_2.
\end{align}
The explicit form of the correction is obtained by plugging the various wave functions and the potential in \eqref{delta-E}, namely
\begin{align}
&\delta E^2= \lambda\,  \frac{b^4\kappa^2}{128} \left\lbrace  b\left( m_1^2+m_2^2+m_3^2 \right) \,+\, \left( |G_{00}|-b \right)\! \sum\limits_{i=1}^2 m_i^2\!
\int\limits_{\!\!\!-1}^1\!\!  \left(1-z_i^2\right)  \left|\Psi_{n_i}^{\alpha_i,\beta_i}(z_i)\right|^2\! dz_i \right. \nonumber\\
&\left. \qquad\qquad\qquad\,\,\,+\, 2b\prod_{i=1}^{2} m_i\! \int\limits_{\!\!\!-1}^1\! z_i \left|\Psi_{n_i}^{\alpha_i,\beta_i}(z_i)\right|^2\! dz_i \,-\, 2bm_3 \sum\limits_{i=1}^2 m_i\!
\int\limits_{\!\!\!-1}^1\! z_i \left|\Psi_{n_i}^{\alpha_i,\beta_i}(z_i)\right|^2\! dz_i  
\right\rbrace .
\end{align}
In short notations it looks like
\begin{align}\label{correction}
&\delta E^2= \lambda\,  \frac{b^4\kappa^2}{128} \left\lbrace  b\left( m_1^2+m_2^2+m_3^2 \right) \,+\, \left( |G_{00}| -b \right)\! \left( m_1^2\, I_1 + m_2^2\, I_2 \right)  \right. \nonumber\\
&\left. \qquad\qquad\qquad\qquad +\, 2b\, m_1m_2\, \tilde{I}_1 \tilde{I}_2 \,-\, 2bm_3\! \left( m_1 \tilde{I}_1 + m_2\, \tilde{I}_2 \right)   
\right\rbrace ,
\end{align}
where the introduced integrals are explicitly calculated as follows
\begin{align}
I_i=&\,\int\limits_{\!\!\!-1}^1\! \left(1-z_i^2\right)  \left|\Psi_{n_i}^{\alpha_i,\beta_i}(z_i)\right|^2 dz_i \nonumber \\
=&\,\frac{(n_i+\alpha_i+\beta_i+1)(n_i+\alpha_i+\beta_i+2)(n_i+\alpha_i+1)(n_i+\beta_i+1)}{(2n_i+\alpha_i+\beta_i+1)
(2n_i+\alpha_i+\beta_i+2)^2(2n_i+\alpha_i+\beta_i+3)} \nonumber \\
&+\dfrac{n_i(n_i+\alpha_i+\beta_i+1)}{(2n_i+\alpha_i+\beta_i+1)^2}\!\left( \frac{n_i+\alpha_i+1}{2n_i+\alpha_i+\beta_i+2}-\frac{n_i+\beta_i}{2n_i+\alpha_i+\beta_i}\right)^2 \nonumber \\	&+\frac{n_i(n_i-1)(n_i+\alpha_i)(n_i+\beta_i)}{(2n_i+\alpha_i+\beta_i+1)(2n_i+\alpha_i+\beta_i)^2(2n_i+\alpha_i+\beta_i-1)}\,,
\end{align}
\begin{align}
\tilde{I}_i&= \int\limits_{-1}^1 z_i\left|\Psi_{n_i}^{\alpha_i,\beta_i}(z_i)\right|^2 dz_i \\
&=\frac{2(n_i+\beta_i)(n_i+\alpha_i+\beta_i)}{(2n_i+\alpha_i+\beta_i+1)(2n_i+\alpha_i+\beta_i)} \,+\,
\frac{2(n_i+1)(n_i+\alpha_i+1)}{(2n_i+\alpha_i+\beta_i+1)(2n_i+\alpha_i+\beta_i+2)}\,.
\nonumber
\end{align}
The expression for the correction to the energy looks very complicated. Therefore, we use the fact that the approximation we work in is for large quantum numbers, say $n_{1,2}\gg 1$. Within this approximation the integrals behave like
\begin{align}
I_i&=\frac{1}{8} + \frac{1}{32}\!\left(2\alpha_i^2+2\beta_i^2-1\right)\!\dfrac{1}{n_i^2}+{\rm O}\!\left(\frac{1}{n_i^3}\right)\!, \\
\tilde{I}_i&=1+\frac{1}{4}\!\left(\beta_i^2-\alpha_i^2\right)\!\dfrac{1}{n_i^2}+{\rm O}\!\left(\frac{1}{n_i^3}\right)\!.
\end{align}
Since $\alpha_i\equiv\ \mid\!l_i-l_3\!\mid$ and $\beta_i\equiv\ \mid\!l_i+l_3\!\mid$, the above integrals look like
\begin{align}
I_i&=\frac{1}{8} + \frac{1}{8}\!\left(l_i^2+l_3^2-\frac{1}{4}\right)\!\dfrac{1}{n_i^2}+{\rm O}\!\left(\frac{1}{n_i^3}\right)\!,\\
\tilde{I}_i&=1+\dfrac{l_il_3}{n_i^2}+{\rm O}\!\left(\frac{1}{n_i^3}\right)\!.
\end{align}
Combining \eqref{Energy} and \eqref{correction} we find the total corrected energy $\tilde{E}=\sqrt{E^2+\delta E^2}$ such as
\begin{align}
\tilde{E}=& \left\lbrace \frac{4\kappa^2}{b^2M^2}\!\left[\frac{b}{4}\sum_{i=1}^2(2n_i + \alpha_i + \beta_i +1)^2 -\left( 2b - \frac{M^2}{\kappa^2} \right)\! l_3^2-\frac{b}{2}\right]	\right. \nonumber \\
& \quad + \lambda\,  \frac{b^4\kappa^2}{128} \left[ b\left( m_1^2+m_2^2+m_3^2\right) \,+\, \left( |G_{00}|-b \right)\! \left( m_1^2\, I_1 + m_2^2\, I_2 \right)  \right. \nonumber\\
&\left. \left. \qquad\qquad\qquad\qquad\qquad   +\,2b\,m_1m_2\, \tilde{I}_1 \tilde{I}_2 \,-\, 2bm_3\! \left( m_1 \tilde{I}_1 + m_2\, \tilde{I}_2 \right)   \right]
\right\rbrace^{1/2}.
\end{align}	
The latter expression can be expanded up to first order in $\lambda$
\begin{align}\label{EnergySeries}
\tilde{E}&=\sqrt{E^2}+ \frac{\delta E^2}{2\sqrt{E^2}}+O(\lambda^2)
\nonumber \\
&=\frac{2\kappa}{bM}\sqrt{ \frac{b}{4}\sum_{i=1}^2(2n_i + \alpha_i + \beta_i +1)^2 -\left( 2b - \frac{M^2}{\kappa^2} \right)\! l_3^2-\frac{b}{2} } \nonumber\\
&\,\,\,\,\,\,+\frac{ b^5\kappa\, M  Y  }{512\sqrt{ \dfrac{b}{4}\sum_{i=1}^2(2n_i + \alpha_i + \beta_i +1)^2 -\left( 2b - \dfrac{M^2}{\kappa^2} \right)\! l_3^2-\dfrac{b}{2} }} \,\lambda \,+\, O(\lambda^2),
\end{align}
where
\begin{align}
Y=&\,\, b\left( m_1^2+m_2^2+m_3^2 \right) \,+\, \left( |G_{00}|-b \right)\! \left( m_1^2\, I_1 + m_2^2\, I_2 \right)	\nonumber\\
&+\, 2b\,m_1m_2\, \tilde{I}_1 \tilde{I}_2   \,-\,2b m_3\, ( m_1 \tilde{I}_1 + m_2\, \tilde{I}_2 ) .	
\end{align}

The anomalous dimension $\Delta$ of the operators in the dual gauge theory is explicitly follows form \eqref{EnergySeries}
\begin{equation}\label{anomalous_dim}
\Delta=\frac{ b^5\kappa\, M Y  }{512\sqrt{ \dfrac{b}{4}\sum\limits_{i=1}^2(2n_i + \alpha_i + \beta_i +1)^2 -\left( 2b - \dfrac{M^2}{\kappa^2} \right)\! l_3^2-\dfrac{b}{2} }} \,\lambda.
\end{equation}

\section{Conclusion}
Our study is focused on the dynamics of pulsating strings in Schr\"odinger background times $T^{1,1}$ conifold with non-zero $B$-field. The Schr\"odinger part of the spacetime was obtained by TsT deformations of the original $AdS_5\times T^{1,1}$ space involving the time direction and one spatial dimension in the internal space. This particular TsT technique is also known as null Melvin twist transformation, which is a special case of the general Drinfel'd-Reshetikhin twist.

Previous studies of pulsating strings in particular backgrounds with external fluxes have been conducted for instance in \cite{Banerjee:2015bia}.   
In order to find pulsating string solutions in $Schr_5\times T^{1,1}$ background, we have employed an appropriate ansatz \eqref{ansatz} for the string configuration.
Considering the bosonic part of the Polyakov string action in conformal gauge we
have found the relevant classical equations of motion \eqref{EOMT}-\eqref{EOMTHETA} together with the Virasoro constraints \eqref{Vir}. The solutions are in therms of Jacobi elliptic functions \eqref{class_sol1}. Furthermore, we managed to explicitly calculate also the classical energy of the pulsating string configuration in terms of the relevant parameters of the problem. 

A great simplification in the semi-classical treatment of the problem came from the fact that the squared Hamiltonian \eqref{Hamiltonian1} of the pulsating strings has the form of a point-particle Hamiltonian. This allowed us to recognize an effective string potential, which was later used to obtain the corrections to the energy via perturbation theory in powers of the small t' Hooft coupling $\lambda$. Holographic AdS/CFT dictionary relates the calculated energy corrections to the anomalous dimensions \eqref{anomalous_dim} of the operators in the dual gauge theory. Thus, the next issue we focused on has been the calculation of corrections to the energy of the bulk theory. The resulting equation \eqref{Schro} for the wave function is a second order Fuchsian differential equation, which is hard to solve analytically. However, since we consider correspondence at quasi-classical level the wave equation can be reduced to simpler expressions applying physically relevant approximation. 

We solved the approximated wave equation and obtained the wave function. Analogously to the case of $AdS_5\times S^5$, the semi-classical quantized energy is given in terms of principal quantum numbers $(l_1,l_2,l_3)$ associated with the variables $\phi_1,\:\phi_2$ and $\psi$ as well as the integer numbers $(n_1,n_2)$ dubbed as quantization conditions. 
The first correction to the energy is obtained by making use of the wave function and potential to the first order in the coupling $\lambda$ \eqref{correction}.  According to the holographic correspondence corrections to the bulk energy determine anomalous dimensions of the dual field theory operators \eqref{anomalous_dim}. As expected, the energy corrections depend on the parameter $\hat{\mu}$ coming from generating background TsT procedure. It measured the deviation from the most studied case of $AdS_5\times T^{1,1}$ \cite{Arnaudov:2010by}. Indeed, the potential for the effective  Hamiltonian \eqref{potential}, compared to \cite{Arnaudov:2010by}, contains deformations not only in the first term but receives a new contribution from the $B$-field. However, if we take a limit $\hat{\mu}\to 0$ the theory should be reduced to holography in $AdS_5\times T^{1,1}$. Indeed, conducting this limit the effective theory in terms of energy correction reduces to the one obtained in \cite{Arnaudov:2010by}.

This study complements the results for non-relativistic holographic correspondence, which is much less investigated compared to the relativistic AdS/CFT correspondence. In this context there are  many ways to proceed. One aspect is to extend this kind of analysis to other backgrounds with non-relativistic field theory duals. Another issue is the identification of (at least of particular class of) field theory operators and thorough investigation of their correlation functions. We hope to address these issues in the near future.

\tocless{\subsection*{Acknowledgments}}

%\paragraph{Acknowledgements}\ \\
R. R. is grateful to Kostya Zarembo for discussions on various issues of holography in
Schrödinger backgrounds. T. V. and M. R are grateful to Prof. G. Djordjevic for the
warm hospitality at the University of Niš, where some of these results have been presented. The work is partially supported by the Program “JINR– Bulgaria” at Bulgarian Nuclear Regulatory Agency. T. V. gratefully acknowledge the support of the Bulgarian national program “Young Scientists and Postdoctoral Research Fellows”.
This work was supported in part by BNSF Grant DN-18/1 and H-28/5, as well as SU Grants 80-10-62/2020 and 80-10-68/2020.

\end{document}